\documentclass{article}
\begin{document}
\title{Theory of the quantification of the redshifts.
\author{Jacques Moret-Bailly 
\footnote{Laboratoire de physique, Universit\'e de Bourgogne, BP 47870, F-21078 Dijon cedex, 
France. email : Jacques.Moret-Bailly@u-bourgogne.fr}}}
\maketitle

\begin{abstract}
The quantification of the redshifts of the lines observed in the spectra of the quasars and the galaxies 
(the redshifts of their lines have some of the values given by the formula $z=a+0.062m$ where $a$ depends on the 
object, and $m$ is an integer), the density of the lines, result from a purely spectroscopic effect. The 
explanation requires only standard spectroscopy, a nearly homogeneous halo of atomic hydrogen, 
and a magnetic field.
\end{abstract}
keywords: plasma, quasars, redshifts.

\section{Introduction}
The  \textquotedblleft Coherent Raman Effect on Incoherent Light\textquotedblright (CREIL) 
\cite{Mor98a,Mor98b,Mor01} explains a lot of properties of the quasars \cite{Mor03}

Creil is a light-matter parametric interaction which has the following properties:

- CREIL redshifts the frequencies of the natural incoherent light, transferring the corresponding energy  to the 
thermal radiation.

- CREIL keeps the widths of the lines, there is no blur of the spectra.

- The relative frequency shift $\Delta \nu/\nu$ is nearly independent on $\nu$.

- As CREIL is a space-coherent process, it does not blur the images.

- CREIL requires a low pressure gas which is not excited, acting as a catalyst; to be active in CREIL, a gas must 
have Raman resonances in the megaHerz range (or lower).

In astrophysics, H$_2^+$ has the convenient resonances. To get them, atomic H needs to be excited by a Lyman 
transition; for low values of the principal quantum number $n$, the fine, hyperfine and Lamb energies are too high 
(gigaHerz range), so that a Zeeman splitting is necessary.
\section{Propagation of light in a halo made of atomic hydrogen.}

Consider a nearly homogeneous halo made of low pressure atomic hydrogen in a low magnetic field.

If the intensity of the nearly thermal spectrum emitted by a strong source does not depend on the frequency, 
supposing that the redshifting power of the gas is constant, the pumping of the Lyman lines is constant, so that 
the constancy of the redshifting power is verified. As absorption and redshift are simultaneous, the width of the 
absorption lines is equal to the redshift, so that the absorptions are spread, weak, invisible.

Suppose that, at some place, the pressure is higher than 100 Pascals, or that the magnetic field disappears locally, 
so that there is no CREIL: the Lyman lines are written into the spectrum.

Then the CREIL appears, spreads the lines, making their absorption invisible. However, if an absorbed, shifted 
and strong (low $n$), Lyman line coincides with an other strong Lyman line, the decrease of CREIL stabilises the 
frequencies, {\it all} absorption lines are written into the spectrum; in particular the absorption of the shifted, 
previously written line is increased. Thus the Lyman absorption patterns become stronger and are linked by 
superpositions of shifted and unshifted frequencies.

\section{Application to the quasars and galaxies.}
The quasars are surrounded by a halo of atomic hydrogen observed by the Lyman absorption and magnetic fields 
are deduced from polarisations. Therefore the previous conditions apply

Using the Rydberg approximation, the coincidence of Lyman $\alpha$ and $\beta$ lines occurs for a shift 
$z_{\alpha\beta}=5/27 = 0.185 = 3 * 0.0617$. For the $\alpha$ and $\gamma$ lines, $z_{\alpha\gamma} = 1/4 =4 * 
0.0625$. By a combination of successive links of patterns, one may obtain $z_f = 2 z_{\alpha\beta} + 
z_{\alpha\gamma} = 0.620$. $z=.062$ may be obtained too. 

These features are observed experimentally \cite{Burbidge,Hewitt,Bell,Comeau}. They apply to the galaxies.

This building of the composed pattern shows that the complexity of the spectrum; the density of lines increases 
with the redshift, an observed property too.
\section{Conclusion.}
There are few hypothesis, which correspond to standard physics and to observations. It is difficult to suppose, 
as the standard theory does, that there are lots of galaxies along the paths of all quasars, and that the redshifts of 
these galaxies obey a single law. The older hypothesis, dark matter or high speed clouds do not work better.

Therefore the CREIL origin of the largest part of the redshifts of the quasars seems proved. Why not the 
remainder to get the \textquotedblleft cosmological redshift  \textquotedblright by a CREIL interaction of the 
light with H$_2^+$ which could provide the 2.7K radiation?

\end{document}